# Disease classification of macular Optical Coherence Tomography scans using deep learning software: validation on independent, multi-centre data

**Author List:**
Kanwal K. Bhatia PhD[1,†], Mark S. Graham PhD[1], Louise Terry PhD[3], Ashley Wood PhD[3], Paris Tranos PhD ICO FRCS[4], Sameer Trikha MBA FRCOphth[1,2], Nicolas Jaccard PhD[1]

[†]Correspondence should be addressed to kanwalb@gmail.com

**Affiliations:**
[1]Visulytix Ltd, Screenworks, 22 Highbury Grove, London, N5 2EF, United Kingdom
[2]King's College Hospital NHS Foundation Trust, London, SE5 9RS, United Kingdom
[3]School of Optometry and Vision Sciences, Cardiff University, Cardiff, CF24 4HQ, United Kingdom
[4]Ophthalmica, Ophthalmology and Microsurgery Institute, 196 Vas. Olgas Avenue & Ploutonos 27, Thessaloniki 546 55, Greece

**Key words:**

Age-related Macular Degeneration, Artificial Intelligence, Clinical Decision Support, Computer Assisted Diagnosis, Deep Learning, Diabetic Macular Edema, Optical Coherence Tomography

**Summary statement:**

Validation of Pegasus-OCT, an artificial intelligence based software for the automated detection of macula disease from OCT scans, is conducted on independent, multi-centre data. 5,588 volumes spanning multiple populations, device manufacturers and acquisition sites were assessed. Pegasus-OCT achieves AUROCs of >98% on AMD, DME and general anomaly detection.

# Abstract

*Purpose:* To evaluate Pegasus-OCT, a clinical decision support software for the identification of features of retinal disease from macula OCT scans, across heterogenous populations involving varying patient demographics, device manufacturers, acquisition sites and operators.

*Methods:* 5,588 normal and anomalous macular OCT volumes (162,721 B-scans), acquired at independent centres in five countries, were processed using the software. Results were evaluated against ground truth provided by the dataset owners.

*Results:* Pegasus-OCT performed with AUROCs of at least 98% for all datasets in the detection of general macular anomalies. For scans of sufficient quality, the AUROCs for general AMD and DME detection were found to be at least 99% and 98%, respectively.

*Conclusions:* The ability of a clinical decision support system to cater for different populations is key to its adoption. Pegasus-OCT was shown to be able to detect AMD, DME and general anomalies in OCT volumes acquired across multiple independent sites with high performance. Its use thus offers substantial promise, with the potential to alleviate the burden of growing demand in eye care services caused by retinal disease.

## Introduction

More than 250 million people globally are currently estimated to be living with moderate to severe visual impairment or blindness[1]. That figure is predicted to double by 2040 due to the projected increase in population growth and ageing[2]. This places a significant burden on healthcare services. Despite there being over 200,000 ophthalmologists worldwide, there is currently a significant shortfall of practitioners in developing countries. Furthermore, although the number of ophthalmologists is increasing in developed countries, the population aged above 60 years is growing at twice the rate of the profession[3]. Prompt and substantive steps are therefore needed to alleviate the current and anticipated deficit of expertise.

Diseases of the macula such as age-related macular degeneration (AMD) and diabetic macular edema (DME) constitute some of the main causes of avoidable sight loss in developed nations. The prevalence of AMD, for instance, was 170 million people in 2016[4], projected to increase to 196 million in 2020, and further to 288 million by 2040[5]. These diseases manifest as anatomical changes to the macula such as the development of drusen, retinal pigment epithelium detachments, and the build up of subretinal and/or intraretinal fluid. The identification of these characteristics is crucial to both detection and to treatment management[6,7,8].

Optical coherence tomography (OCT) imaging is currently the most frequently used imaging modality in the evaluation of macular disease, with over 30 million scans taken per year[9]. For the diagnosis of AMD, a 70-fold increase in the use of OCT has been recorded since its introduction[10] and it has additionally become a critical tool for the baseline retinal evaluation needed to guide administration of therapy[11,12].

OCT imaging is now widely regarded as the gold standard for guiding the diagnosis and treatment of AMD and DME. However, the resulting rapid expansion in imaging data generated has not been matched by the availability of trained experts to interpret these scans accurately. Computer-aided diagnosis (CAD) has the potential to alleviate the human and economic burden of disease detection. Recently, CAD systems that utilise Artificial Intelligence (AI) in particular have gained prominence, for instance in the detection of Diabetic Retinopathy (DR) from fundus photographs[13,14]. However, the use of AI CAD systems for the analysis of OCT image data has so far been less explored.

The applicability of AI in ophthalmic diagnosis has become particularly propitious since the development of a method known as deep learning[15,16]. Deep learning uses convolutional neural networks to process grid-based data such as images, and has been observed to be particularly successful in automated imaging diagnostics[17]. A deep learning algorithm aims to mimic the way the human expertise is developed for the same task by discovering image features that distinguish normal from pathological classes. These features are then used to classify new images when the algorithm is applied. However, a potential risk is that poor or unexpected performance may occur if the data used to train the algorithm inadequately captures the diversity of patients and images present in real world scenarios[18]. While humans can generalise knowledge learned on one set of images to others which may have very different appearances, deep learning algorithms are prone to overfitting or bias resulting from limited variation in training data[19]. A thorough evaluation of any algorithm is therefore necessary before adoption in clinical practice.

In the interpretation of ophthalmic images using deep learning, the largest body of work to date has been in the detection of DR from fundus photographs[20,21,22,23]. The use of these methods in clinical practice has also been shown in retrospective trials to provide significant reduction in the manual burden and costs of screening for DR[24,25]. Although research in OCT analysis is more incipient, papers released in recent years have shown promising results[26,27,28,29,30].

The largest scale prior investigations[28,29,30] evaluate their algorithms on 1000-2151 OCT volumes. However, common to all the above algorithms is a concern that results are presented on evaluation sets which are from the same cohorts as those that the algorithms were trained on. Questions therefore remain as to how these algorithms would perform on examples from cohorts external to the training set[18]. To date, none of the algorithms in the papers described above have been made available by their authors for external evaluation.

In the case of OCT it has been shown that, even in the absence of disease, variations in ethnicity, age and gender can result in statistically significant differences in the appearance of retinal structures[31,32]. In addition, imaging variation due to the acquisition device needs to be accounted for.

This paper provides an evaluation of Pegasus-OCT, a commercial AI based system for the detection of retinal diseases from OCT images of the macula. In contrast to prior work in the field, the evaluation is conducted on independent datasets spanning a range of acquisition sites and scanner types, on a total of 5,588 OCT cube volumes (162,721 B-scans). To the best of our knowledge, it represents the first evaluation of deep learning based algorithms for OCT image interpretation on data originating from populations that have not in any way been utilised in creating the algorithms.

## Methods

**Automated Disease Detection**

Pegasus-OCT v1.0 is a clinical decision support system for detecting disease from OCT scans of the macula developed by, and available from Visulytix Ltd (Visulytix, London, UK), currently for investigational purposes. Using deep learning technologies, the software identifies images showing anomalous features in the macula which could be indicative of disease. In addition, specific features indicative of Dry AMD, Wet AMD and DME are also identified. This system produces four independent classifications in line with the guidelines outlined by the National Institute of Clinical Excellence[33]:

1. General anomaly (e.g. presence of signs of unspecified pathology or anomalous features)

2. 'Dry' AMD incorporating signs of early and atrophic AMD (e.g. presence of drusen, pigmentary abnormalities, uncomplicated pigment epithelial detachments or geographic atrophy)

3. 'Wet' AMD incorporating active and inactive neovascular AMD (e.g. presence of intra- or sub-retinal fluid, along with fibrovascular/serous PED and/or signs of Dry AMD)

4. DME (e.g. presence of hyper reflective foci, retinal thickening, and intra- and /or sub-retinal fluid)

Pegasus-OCT reports a score for each of these categories, showing the likelihood of features of these pathologies being present. Features may be common to more than one category; for instance, features of Dry AMD (such as drusen) are often also visible in patients with Wet AMD. The detection of drusen may therefore result in a high score for both of these categories if other features of Wet AMD (such as fluid) are also present. An additional classification of General AMD, obtained by pooling Wet and Dry predictions, is therefore also presented in this paper.

**Deep Learning Models**

The four classifiers used are all binary classifiers which utilise the same VGG16 convolutional neural network model[34]. Details of the network architecture and training details are given in Supplemental Digital Content 1. These models have been trained on OCT data acquired from different manufacturers and using a range of imaging protocols. Before training, all images are normalised to have the same size and intensity range. In addition, to increase generalisability, significant image augmentation is used during training. At each training epoch, images are read in with a random rotation, translation, zoom and brightness enhancements applied. This means that the model never sees the same image twice during training and serves to enlarge the range of training data. Finally, to prevent overfitting to the training data, early stopping is used[19]. These extensions enable the application of the classifiers to images acquired from different scanners and with differing B-scan resolutions.

**Image Quality Assessment**

An automated strategy for disease detection requires images of sufficient quality to reliably visualise retinal anatomy. Poor contrast, extensive noise, or artifacts may conceal the presence of real pathology or may be confused for abnormalities in otherwise normal scans. We therefore also verify the quality of each B-scan in terms of suitability for automated disease detection. This is done via a deep learning method which classifies scans as being either 'gradable' or 'ungradable'. An overall quality rating is then reported for each dataset, reflecting the percentage of gradable B-scans in the dataset.

**Performance Evaluation**

In this paper, the results of Pegasus-OCT in classifying each of the four categories are evaluated on five external datasets. These datasets were acquired independently of the Pegasus-OCT manufacturer as well as independently of each other. Local ethical committee approval was obtained for each dataset where required. The datasets span geographic locations and multiple acquisition sites, and were acquired with a range of protocols and scanners. A summary of the datasets used in the evaluation of Pegasus-OCT is shown in Table 1. Where available, demographic information is given in Supplemental Digital Content 2. An example image from each dataset can be seen in Figure 1.

Images from these datasets were run through the Pegasus-OCT software with the operator masked to ground truth diagnoses. The results of Pegasus-OCT were then compared to the provided ground truth at a whole volume level. However, for dataset A1, whole macular volumes were not made available for assessment. Instead, ground truth is provided for each individual B-scan in the dataset and so the results are assessed at a B-scan level in this case only.

All classifiers are run on all images which report a probability of anomaly or pathology for each of the four cases. The classification of General Anomaly is based solely on the probability of this classifier exceeding a specified threshold, where the same threshold is applied across all datasets. For the pathology classes, if only one classifier yields a probability above the relevant threshold, the subject is classified with that pathology. However there are cases in which two or more classifiers report probabilities above the required threshold. In the case of Wet and Dry AMD being present, the volume will always be classified as Wet AMD. Where both AMD and DME are present, the classification will be assigned to that with the highest probability. It should be noted that while this allows quantitative evaluation of the algorithms, having all probabilities available to a user in a clinical setting may be helpful in the assessment of complex cases.

The performance of Pegasus-OCT was evaluated by calculating the area under the curve of the receiver operating characteristic (AUROC) curves as in comparable work[28,29,30]. The AUROC is a measure of the diagnostic ability of a predictor in a binary classification. It gives the probability that the classifier will rank a randomly chosen pathological instance higher than a randomly chosen instance without that pathology.

**Datasets from Public Sources**

Dataset A1[29] consists of 108,309 macula OCT images from 4,686 patients acquired on Heidelberg Spectralis OCT scanners. These images consist of individually-labelled B-scans only; whole volumes were not available. Subjects were taken from retrospective cohorts of adult patients from the Shiley Eye Institute of the University of California San Diego, the California Retinal Research Foundation, Medical Center Ophthalmology Associates, the Shanghai First People's Hospital, and Beijing Tongren Eye Center between July 1, 2013 and March 1, 2017. All OCT imaging was performed as part of patients' routine clinical care. There were no exclusion criteria based on age, gender, or race. Further details of the demographics of the cohort can be found in Supplemental Digital Content 2. It should be noted that the mean ages of normal and pathological cases differ significantly and this should be considered when assessing results. Local electronic medical record databases for diagnoses were searched and each 2D B-scan image was then graded by a tier of graders including four ophthalmologists and two senior retinal specialists. This dataset was used by the authors to detect patients requiring urgent referral and observational cases. However, due to the difference in labelling between this dataset and the Pegasus-OCT pathology classification, this dataset is only used for the evaluation of General Anomalies.

Dataset A2[35] consists of 384 cases (AMD and Normal Controls). Volumetric scans with non-unique acquisition protocols were acquired from four SD-OCT clinics (Devers Eye Institute, Duke Eye Center, Emory Eye Center and National Eye Institute, USA), on Bioptigen OCT scanners. Subjects were recruited from the AREDS2 Clinical Trial NCT00734487. The A2A SDOCT study recruited AMD subjects from the AREDS 2 study population at 4 AREDS 2 Study Centers. Controls were recruited from Duke University Eye Center and Emory University. Subjects included were AMD subjects and controls, men and women between the ages of 50 and 85 years. The AMD subjects had macular status ranges from large drusen in both eyes or large drusen in one eye and advanced AMD (neovascular AMD or geographic atrophy) in the fellow eye. Subjects were excluded only if the ocular media was not clear enough to allow good fundus photography.

Dataset A3[36] consists of 148 cases with a distribution of 50 normal controls, 48 Dry AMD, and 50 DME, acquired at Noor Eye Hospital in Tehran on Heidelberg SD-OCT imaging systems. These data contains varying number of both A-scans (512 or 768) and B-scans (range 19-61). No further demographic information is available.

**Datasets from Private Sources**

Dataset B1 consists of 25 DME, 25 Dry AMD, 25 Wet AMD, and 25 normal controls, imaged on Heidelberg SD-OCT devices. Anonymisation was conducted at source. Subjects from each category were selected at random from routine clinical care from the Ophthalmica clinic in Thessaloniki, Greece.

Dataset B2 consists of 135 eyes acquired on a Topcon 3D-OCT 1000 device (B2a) and the same 135 eyes acquired on a Zeiss Cirrus HD-OCT 5000 device (B2b) at the School of Optometry and Vision Sciences, Cardiff University. Macular volumes from each device comprised 512x128 scans. Ground truth was established using a modified AREDS classification system, incorporating OCT images. Colour fundus photos and OCT images were presented in conjunction, and classified by two experienced clinicians. Any disagreements were classified by a third observer and discussed by all three to reach a final classification. Demographic data for this cohort are given in Supplemental Digital Content 2.

## Results

Results of the evaluation of Pegasus-OCT on all datasets are shown in Table 2. The Quality Rating is an automated assessment of the percentage of B-scans in each dataset deemed to be of sufficient quality for automated grading. The same threshold for quality was used across all datasets.

For the purposes of this paper, evaluation was conducted on all images regardless of quality. Accuracy, sensitivity, specificity and AUROC results were reported at a whole macular volume level, using ground truth provided by the dataset owners. For dataset A1 only, volume-level labels were not available. However, all 108,309 B-scans were individually labelled by the providers, and so the metrics are evaluated on a B-scan level in this case. All accuracy, sensitivity and specificity results have been reported at the same threshold on the ROC curve for all datasets. Actual ROC curves can be found in Supplemental Digital Content 3.

In distinguishing between normal and abnormal scans, Pegasus-OCT performs at an AUROC of over 98% for every dataset. In datasets with quality ratings of above 50%, the minimum AUROCs obtained for the detection of general AMD and DME were found to be 99% and 98%, respectively. Lower performance was observed when the image quality of the B-scans was deemed to be insufficient for automated grading. It should be noted that ground truth for datasets such as B2 were obtained using clinical information in addition to the OCT volume presented.

## Discussion

Clinical decision support systems for the analysis of macula OCT scans offer substantial promise. Such systems can be agile and their performance adjusted for the specific integration in workflow. For example, in a screening context, specificity could be increased to avoid unnecessary false positives which can burden secondary care[25]. Similarly they have the potential to be deployed at scale and therefore at low cost, thus benefiting healthcare providers[24].

Crucial to the translation of algorithms from theory to the clinic is the idea of generalisability. An algorithm needs to perform well on data acquired from a variety of populations, devices, operators and protocols. The power of deep learning algorithms lies in their flexibility and capacity to encode very large and complex amounts of information. However, this also makes deep learning algorithms particularly vulnerable to bias stemming from the data they are trained on (overfitting), which can result in catastrophic failures when applied in the real world[18,37]. Despite this, even the most notable publications have only evaluated their algorithms on subjects from the same populations as those on which they were trained. While providing promising preliminary results, the extrapolation of the stated performance of these algorithms to more diverse data sources cannot - and must not - be assumed.

In contrast, this paper conducts an evaluation of Pegasus-OCT on purely external datasets, which were not used in the creation of the algorithms. It represents validation of an AI system on the largest and most diverse OCT data to date, to the best of our knowledge, with data acquired from multiple sites in five different countries analysed. The heterogeneity of the datasets used aims to address any potential selection biases. A comparison of the evaluation datasets used in the work presented here and in pertinent prior work is shown in Table 3.

The results show promising performance of Pegasus-OCT. Evaluation has been conducted on large datasets from multiethnic populations, without any specific optimisation. Accuracy, sensitivity and specificity analysis was conducted at a common threshold on the ROC curve across all datasets. Improved performance is likely to be obtained when the cut-off point on the ROC curve is chosen specifically to optimise the characteristics of a particular clinical site.

These results presented in this paper are comparable to those of prior related work in Table 3. In the separation of Normal and AMD patients, an AUROC of 93.8% for classification of macula cube volumes and 97.5% when multiple volumes from the same patient were combined for an overall diagnosis was reported[28]. Dataset A1 was used to distinguish between urgent referrals and observation cases with a resulting AUROC of 99.9%[29]. Most recent work[30] reported

AUROCs for disease detection of 99.5% and individual pathology detection ranges between 96.6% to 100%. The key difference is that these papers use between 80-99% of their cohort data to train their algorithms, with the evaluation conducted on the remaining subjects in the cohort.

While the results presented in this paper have been evaluated on data from a range of external sources, the distribution of scanner manufacturer used is still unbalanced. The majority of data have been evaluated on Heidelberg scanners which also show the best performance. However, these datasets were also assessed to have the highest quality scans. Further work is needed to disambiguate the influence of scanner manufacturer and image quality in the performance of Pegasus-OCT.

The study presented evaluated defined datasets including normal, wet, dry AMD and DME. Other disease types, such as Macular/Lamellar Hole, Epiretinal Membrane (ERM) were not explicitly labelled in the datasets and so were not independently assessed in this version of the platform. However signs of these may have been present in images of other pathology (such as Dry AMD with ERM).

As the datasets were mainly taken from trials or secondary care, the prevalence of the diseases evaluated in this paper are likely to be lower and milder in a screening setting. However, by adjusting sensitivity and specificity thresholds, performance can be tailored to suit the intended use of the software. This opens up exciting and transformative opportunities to utilise OCT in non-specialist settings, with subsequent work required to determine the health economic effects of such system implementation. Furthermore, the use of such commercially available platforms has the potential to assist clinicians in managing the exponential demand in eye care services caused by retinal disease.

**Acknowledgements**

Prof. Tom H. Margrain and Prof. Rachel V. North at the School of Optometry and Vision Sciences, Cardiff University, for their contribution to the acquisition and classification of Dataset B2.

**Funding:** This study was funded by Visulytix Ltd.

*Table 1: Independent evaluation datasets used in this paper*

| Name | Acquisition device manufacturer(s) | Countries | Number of acquisition Sites | Number of eyes | Source |
|---|---|---|---|---|---|
| **A1** | Heidelberg SD-OCT | USA<br>China | 3<br>2 | 4,686 | Public[29] |
| **A2** | Bioptigen SD-OCT | USA | 4 | 384 | Public[35] |
| **A3** | Heidelberg SD-OCT | Iran | 1 | 148 | Public[36] |
| **B1** | Heidelberg SD-OCT | Greece | 1 | 100 | Private |
| **B2** | a) Topcon 3D-OCT 1000<br>b) Zeiss Cirrus HD-OCT 5000 | UK | 1 | 135<br><br>135 | Private |
| **Total number** | 4 unique manufacturers | 5 | 12 | 5,588 | |

*None of the images in these datasets were used in the development of Pegasus-OCT.*

*Table 2: Results of Pegasus-OCT evaluation showing Area under Receiver Operating Characteristic (ROC) curve, Accuracy (Acc), sensitivity (Se) and specificity (Sp) of detection. Reported figures correspond to evaluation on a whole OCT volume level, unless otherwise stated.*

| | Number of OCT volumes (slices) | Performance | | | | | Quality Rating (%) |
|---|---|---|---|---|---|---|---|
| | | General Anomaly | General AMD | Dry AMD | Wet AMD | DME | |
| **A1** (Heidelberg SD-OCT) Shiley Eye Institute of the UCDS (USA), California Retinal Research Foundation (USA), Medical Center Ophthalmology Associates, Shanghai First People's Hospital, (China) Beijing Tongren Eye Center (China) | 4,686 (108,309) | ROC: 0.99* Acc: 0.96 Se: 0.94 Sp: 0.98 | | | | | 94 |
| **A2** (Bioptigen SD-OCT) Devers Eye Institute, Duke Eye Center, Emory Eye Center and National Eye Institute (USA) | 384 (38,382) | ROC: 0.98 Acc: 0.91 Se: 0.98 Sp: 0.71 | ROC: 0.89 Acc: 0.89 Se: 0.93 Sp: 0.79 | | | | 14 |
| **A3** (Heidelberg SD-OCT) Noor Eye Hospital, Tehran (Iran) | 148 (2,960) | ROC: 0.99 Acc: 0.94 Se: 0.92 Sp: 0.98 | ROC: 0.99 Acc: 0.95 Se: 0.96 Sp: 0.95 | ROC: 0.98 Acc: 0.94 Se: 0.96 Sp: 0.93 | | ROC: 0.98 Acc: 0.93 Se: 0.96 Sp: 0.91 | 89 |
| **B1** (Heidelberg SD-OCT) Ophthalmica Ophthalmology | 100 (12,800) | ROC: 1.00 Acc: 0.98 Se: 1.00 Sp: 0.92 | R: 0.999 Acc: 0.96 Se: 1.00 Sp: 0.92 | ROC: 0.99 Acc: 0.94 Se: 1.00 Sp: 0.92 | ROC: 0.98 Acc: 0.96 Se: 0.96 Sp: 0.96 | ROC: 0.998 Acc: 0.96 Se: 0.96 Sp: 0.96 | 89 |
| and Microsurgery Institute (Greece) | | | | | | | |
| **B2a** (Topcon 3D-OCT 1000) School of Optometry and Vision Sciences, Cardiff University (UK) | 135 (17,280) | ROC: 0.99 Acc: 0.96 Sens: 0.95 Spec: 0.96 | ROC: 0.98 Acc: 0.87 Se: 0.99 Sp: 0.64 | ROC: 0.85 Acc: 0.71 Se: 0.86 Sp: 0.66 | ROC: 0.86 Acc: 0.87 Se: 0.89 Sp: 0.87 | | 49 |
| **B2b** (Zeiss Cirrus HD-OCT 5000) School of Optometry and Vision Sciences, Cardiff University (UK) | 135 (17,280) | ROC: 0.99 Acc: 0.95 Se: 0.95 Sp: 0.94 | ROC: 0.99 Acc: 0.95 Se: 0.95 Sp: 0.94 | ROC: 0.83 Acc: 0.76 Se: 0.80 Sp: 0.74 | ROC: 0.93 Acc: 0.85 Se: 0.91 Sp: 0.82 | | 87 |

\* AUROC evaluated at B-scan level, not whole volume level

*Table 3: Evaluation datasets used in deep learning methods for OCT classification*

| Publication | Testing set acquisition sites and scanners | Testing set size (3D OCT volumes unless otherwise stated) | Testing set data sources used in training? (% of cohort data used for training) |
|---|---|---|---|
| De Fauw et al. 2018[30] | ● Moorfields, UK (Heidelberg)<br>● Moorfields, UK (Topcon) | 1,113 | ✔ All<br>(93% of all data) |
| Kermany et al. 2018[29] | ● Shiley Eye Institute of the UCDS, USA<br>● California Retina Research Foundation, USA<br>● Medical Center Ophthalmology Associates, USA<br>● Shanghai First People's Hospital, China<br>● Beijing Tongren Eye Center, China<br>(All Heidelberg) | 1,000<br>(2D slices) | ✔ All<br>(99% of all data) |
| Lee et al. 2017[28] | U. Washington (Heidelberg) | 2,151 | ✔ All<br>(80% of all data) |
| This paper | ● Shiley Eye Institute of the UCDS, USA<br>● California Retina Research, Foundation, USA<br>● Medical Center Ophthalmology Associates, USA<br>● Shanghai First People's Hospital, China<br>● Beijing Tongren Eye Center, China<br>(All Heidelberg)<br>● Ophthalmica Institute, Greece (Heidelberg) | 5,588 | ✘ None |
| | ● Noor University Hospital, Tehran, Iran (Heidelberg)<br>● Devers Eye Institute, USA<br>● Duke Eye Center, USA<br>● Emory Eye Center, USA<br>● National Eye Institute, USA<br>(All Bioptigen)<br>● Cardiff University, UK (Zeiss)<br>● Cardiff University, UK (Topcon) | | |

**Figure 1:** Example images from each dataset evaluated

(a) A1

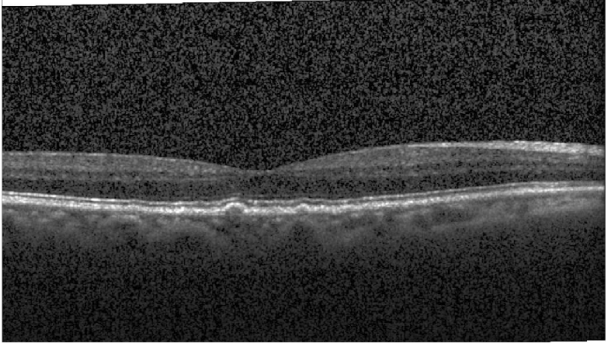

(b) A2

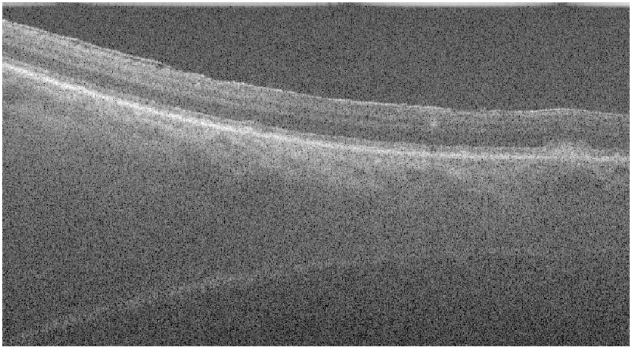

(c) A3

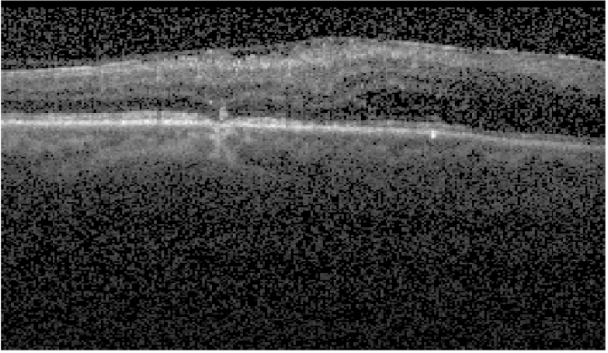

(d) B1

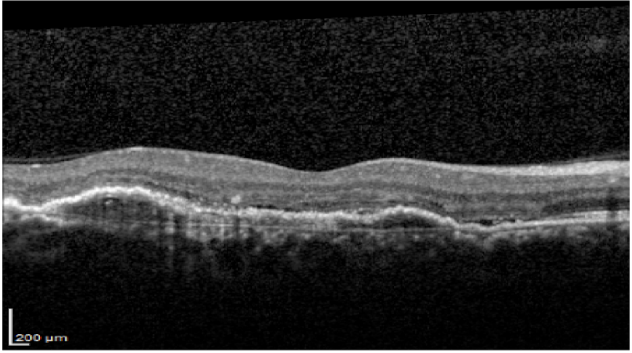

(e) B2a

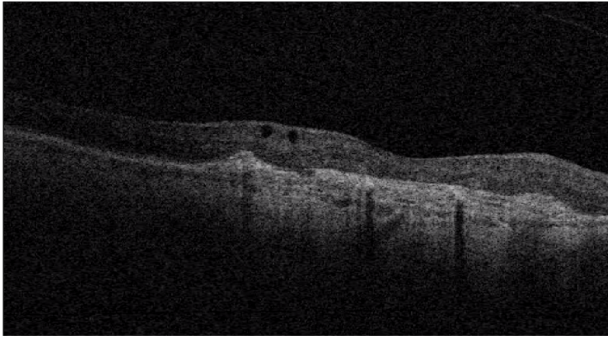

(f) B2b

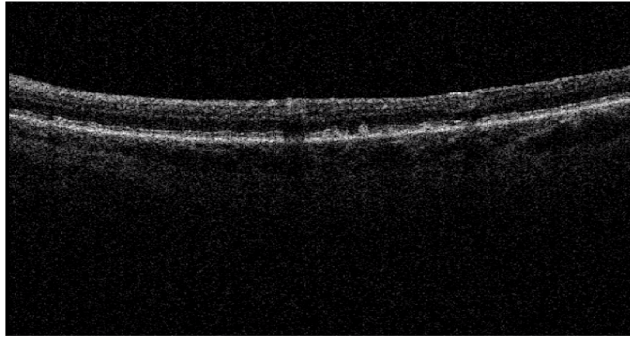